\documentclass[12pt]{article}
\usepackage{graphicx}
\usepackage{amsfonts}

\def\cA{{\cal A}}

\def\cC{{\cal C}}

\def\cM{{\cal M}}
\def\qq{\overline{q}q}
\def\uu{\overline{u}u}
\def\dd{\overline{d}d}
\def\be{\begin{equation}}
\def\ee{\end{equation}}
\newcommand{\GS}[1]{#1\!\!\!\!\!\not~}

\newcommand{\tr}{\mbox{tr}}

\begin{document}
{\large
\hfill\vbox{\hbox{DCPT/06/154}
            \hbox{IPPP~/06/~77}}}
\nopagebreak
\vspace{3.0cm}

\begin{center}
{\LARGE
{\bf $\qq$ condensate for light quarks \\[3mm]beyond the chiral limit}}\\

\vspace{15mm}

{\large{R. Williams$^{1}$, C. S. Fischer$^{1,2}$ and M. R. Pennington$^{1}$}}\\

\vspace{7mm}

{$^{1}$Institute for Particle Physics Phenomenology, Durham University,\\ Durham  DH1 3LE, UK}\\
{$^{2}$ Institut f\"ur Kernphysik, Technische Universit\"at Darmstadt, \\Schlossgartenstra\ss e 9, 64289 Darmstadt, Germany}
\end{center}

\vspace{2cm}

\centerline{\bf Abstract}
\vspace{3mm}
\small

We determine the $\qq$ condensate for quark masses from zero up to that of the strange quark within a phenomenologically successful modelling of continuum QCD
by solving the quark Schwinger-Dyson equation. The existence of multiple solutions to this equation is the key to an accurate and reliable extraction of this condensate using the operator product expansion. 
We explain why alternative definitions fail to give the physical condensate.

\normalsize

\newpage
\parskip=2mm
\baselineskip=5.5mm

\section{Introduction}
Low energy  hadron dynamics is controlled by the breaking of chiral symmetry,  directly reflecting the non-trivial structure of the QCD vacuum. This vacuum is dominated by  long range correlations between quarks and antiquarks. It is the scale of this $\qq$ condensate that determines the mass of constituent quarks and hence the masses of all light hadrons.
Recently  experiments involving $\pi\pi$ interactions at and close to threshold have allowed the $\uu$, $\dd$ condensates to be extracted and these confirm a size of $\sim\,-(235 \,{\rm MeV})^3$ anticipated from phenomenology~\cite{Pennington:2005be}.  How does this value depend on the current mass of the quarks within QCD?
  
The Schwinger-Dyson equations are the natural tool for investigating strong coupling QCD in the continuum. To study the momentum dependence of the quark propagator one needs to know the product of the quark-gluon vertex and the gluon propagator. Models for these have been developed over the last decade with remarkable phenomenological success. The models have sufficient interaction strength below 1 GeV 
that the vacuum becomes non-trivial and a mass gap is created~\cite{Maris:1997hd,Maris:1998hc}. Consequently,
quarks with zero current mass have a dynamically generated mass. Indeed, the momentum dependence of the quark propagator for any current mass is determined. The aim of this paper is to extract
from this dependence
 the behaviour of the value of the $\qq$ condensate beyond the limit of vanishing quark mass.

The interest in the value of such a condensate arises in the context of QCD sum-rules. There the Operator Product Expansion (OPE) is used to approximate the short distance behaviour of QCD. In studying currents like that of $\,\overline{q_i} \gamma^{\mu} (\gamma_5) q_j$, with $q_i=s$ and $q_j=u,d$, the vacuum expectation values of $\uu$, $\dd$ and $\overline{s}s$ operators naturally arise~\cite{Jamin:2006tj,Jamin:2002ev,Jamin:2001fw}. The value of the $\qq$ condensate for the $u$ and $d$ quarks now well determined to be $-(235 \pm 15\, {\rm MeV})^3$ by experiment~\cite{Colangelo:2001sp} is the result in the chiral limit. However, in the OPE it is the value of the condensates away from this limit that actually enters. Since the current masses of the $u$ and $d$ quarks are only a few MeV, the resulting condensate is expected to be close to its value in the chiral limit, but how close? For the first 20 years of QCD sum-rules
their accuracy was never sufficient for it to matter whether this
difference was even a 10\% effect.
This equally applies to the estimate by Shifman, Vainshtein and Zakharov~\cite{Shifman:1978bx,Shifman:1978by} that
the $\overline{s}s$ condensate was $(0.8\,\pm\,0.3)$ of the $\uu$ and $\dd$ values. It is the greater precision brought about  by the studies of Refs.~\cite{Jamin:2006tj,Maltman:2001jx,Maltman:2002sb}, for instance, that motivate the need to learn about how the $\qq$ condensate depends on the current quark mass.
Our aim here is to illustrate a method for determining this dependence.

For light quarks, $u$, $d$ and $s$, studying the 
Schwinger-Dyson equation for the fermion propagator in the continuum is essential (see for instance~\cite{Fischer:2006ub}), until computation with large lattice volumes with small spacing  become competitive.
\section{Schwinger-Dyson Equations}
\begin{figure}[b]
\vspace{1mm}
 \begin{center}  
       \includegraphics*[width=0.95\columnwidth]{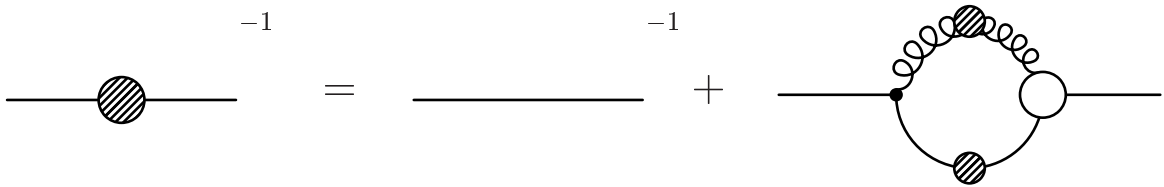}
	\caption{Schwinger-Dyson equation for the quark propagator}\label{quark:sde}
\end{center}  
  \end{figure}
\begin{figure}[ht]
\vspace{1mm}
 \begin{center}  
       \includegraphics*[width=0.9\columnwidth]{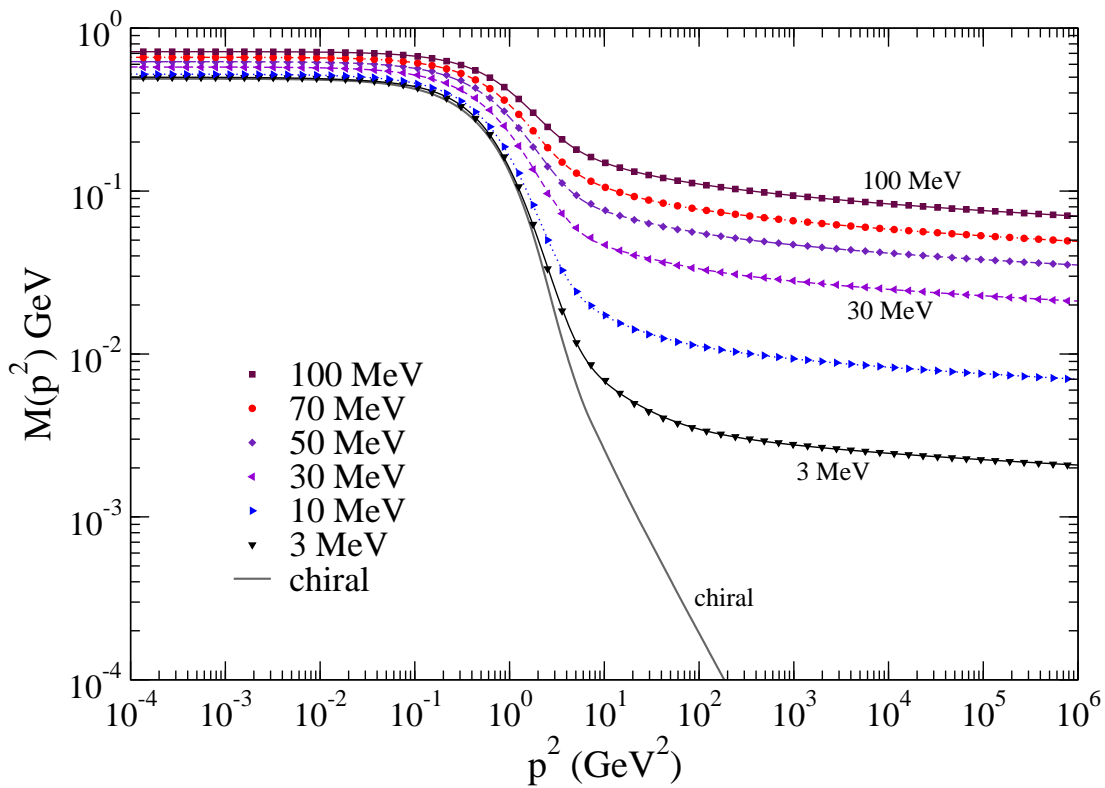}
	\caption{Euclidean mass functions for different current masses, specified at $\mu = 19$ GeV as labelled. The plot illustrates how on a log-log plot the behaviour dramatically changes between a current mass of 0 and 3~MeV.
These results are essentially the same as found by Maris and Roberts~\protect{\cite{Maris:1998hc}}.}\label{masses}
\end{center}  
  \end{figure}

 
Our aim is to calculate the mass function of the quark propagator for a range of current masses. The starting point is the renormalized Schwinger-Dyson equation for the quark propagator as depicted in Fig.~\ref{quark:sde}:
\begin{eqnarray}\label{cond:eqn:quarksde1}	
\vspace{2mm}
	S_F^{-1}(p) &=&  Z_2\left[ S_F^{(0)}(p) \right]^{-1}-C_F \frac{\tilde{Z}_1\,Z_2}{\tilde{Z}_3}\frac{g^2}{\left( 2\pi \right)^4}\int d^4k \cdots\nonumber\\[2mm]
	&&\hspace{35pt}\times\gamma_\mu\, S_F(k)\,\Gamma_\nu(k,p)\,D_{\mu\nu}(p-k)\;.
\vspace{2mm}
\end{eqnarray}
In the Landau gauge we note that $\tilde{Z}_1 = 1$. The inverse propagator $S_F^{-1}(p)$ is specified by two scalar functions $\cA$ and $\cM$:
\begin{equation}\label{fermion:eqn:AB}
	S_F^{-1}(p)\,=\,\cA(p^2)\left(\GS{p} +\;M(p^2)\right)\; .
\end{equation}
While $\cA$ is also a function of the renormalisation point $\mu$ and so strictly $\cA(p^2,\mu^2)$, the quark
mass function $M(p^2)$ is renormalisation group invariant.
Of course, Eq.~(\ref{cond:eqn:quarksde1}) involves not just the quark propagator, $S_F$, but the full quark-gluon vertex, $\Gamma^{\nu}$, and the gluon propagator $D^{\mu\nu}$. 
Rather than solving for the coupled quark, ghost and gluon system, one may employ some suitable ansatz for the coupling  and interaction in Eq.~(\ref{cond:eqn:quarksde1}) which has sufficient integrated strength in the infrared to achieve dynamical mass generation. There have been many suggestions  for this in the literature~\cite{Maris:1997hd,Maris:1998hc} which have been extensively studied. Following the lead of Maris {\it et al.}~\cite{Maris:1997hd,Maris:1998hc}, we will employ an ansatz for $g^2D_{\mu\nu}(p-k)$ which has been shown to be consistent with studies of bound state mesons, and consider other modellings elsewhere. Since this simple model assumes a rainbow vertex truncation, the solutions are not multiplicatively renormalisable and so depend on the chosen renormalisation point. The renormalisation scheme is one of modified momentum subtraction at some point $\mu$. This we take to be at $19$ GeV to compare with earlier studies
\cite{Chang:2006bm}.
 We will later evolve $\mu$ to the more common 2 GeV scale in the $\overline{MS}$-scheme. We use:
\begin{eqnarray}
	\frac{g^2}{4\pi}\frac{Z_2}{\tilde{Z}_3}D_{\mu\nu}(q)\rightarrow \alpha\left(q^2\right) D^{(0)}_{\mu\nu}(q)
\end{eqnarray}

\noindent where the coupling is described by ~\cite{Maris:1997hd,Maris:1998hc}:
\begin{eqnarray}
&&\nonumber\\ \label{eqn:modelparam}
	\alpha\left( q^2 \right) &=& \frac{\pi}{\omega^6}\,D\, q^4\, \exp(-q^2/\omega^2)\nonumber\\[2.mm]
	&+&\frac{2\pi \gamma_m}{\log\left( \tau+\left(1+q^2/\Lambda_{QCD}^2 \right)^2\right)}\nonumber\\[0.mm]
	&&\times \left[ 1-\exp\left(-q^2/\left[ 4m_t^2 \right]\right) \right]\;,
\end{eqnarray}
with 
\begin{eqnarray}
\nonumber m_t&=& 0.5\;{\rm GeV}\qquad\qquad,\qquad\tau\;=\;\textrm{e}^2-1\qquad\;,\\
\nonumber \gamma_m&=&12/(33-2N_f)\,\quad,\quad\Lambda_{QCD}\;=\;0.234\,{\rm GeV}\; .
\end{eqnarray}
We work with both the $N_f=0, 4$ limits. The precise value of $\Lambda_{QCD}$ is irrelevant for our current study. The parameters $\omega$ and $D$ control the infrared behaviour of the coupling $\alpha(q^2)$ for momenta less than $\Lambda_{QCD}$, or strictly speaking $m_t$. The pion decay constant correlates these parameters. Consistent with this we choose $\omega=0.4$ GeV, $D=0.933$ GeV$^2$ as a typical parameter set.
Solutions are obtained by solving the coupled system of fermion equations for $\cA$ and $\cM$ of Eq.~(\ref{fermion:eqn:AB}), which we may write symbolically as:
\begin{eqnarray}
	\cA(p^2,\mu)&=&Z_2(\mu,\Lambda) - \Sigma_D\left(p,\Lambda\right)\; ,\nonumber\\[-1mm]
&& \\[-1.mm]
	\cM(p^2)\cA(p^2,\mu)&=&Z_2(\mu,\Lambda) Z_m m_R(\mu) + \Sigma_S\left(p,\Lambda\right)\;.\nonumber
\end{eqnarray}
The $\Sigma_S$ and $\Sigma_D$ correspond to the 
scalar and spinor projections of the integral in Eq.~(\ref{cond:eqn:quarksde1}). For massive quarks we obtain the solution $M$ (later called $M^+$) by eliminating the renormalisation factors $Z_2$, $Z_m$ via:
\begin{eqnarray}
	Z_2(\mu,\Lambda)&=& 1+\Sigma_D\left( \mu,\Lambda \right)\; ,\nonumber\\[-2.mm]
  && \\[-2.mm]
	Z_m(\mu,\Lambda)&=& \frac{1}{Z_2(\mu,\Lambda)}-\frac{\Sigma_S\left(\mu,\Lambda \right)}{Z_2(\mu,\Lambda) m_R\left(\mu\right)}\;.\nonumber
\end{eqnarray}
The momentum dependence for different values of $m_R$ are shown in Fig.~\ref{masses}. Our purpose is to define the value of the $\qq$ condensate for each of these.

At very large momenta the tail of the mass function is described by the operator product expansion so that
\begin{eqnarray}\label{cond:eqn:opefit}
\vspace{0.5mm}
	M(p^2)_{asym} &=&  \overline{m}\left[ \log\left( p^2/\Lambda_1^{\,2} \right] \right)^{-\gamma_m}\nonumber\\[2.mm]
  &+&\frac{2\pi^2}{3}\frac{\cal C}{p^2}\left[ \frac{1}{2}\log\left( p^2/\Lambda_2^{\,2} \right) \right]^{\gamma_m-1}\;.
\vspace{0.5mm}
\end{eqnarray}
where the first term corresponds to the explicit mass in the Lagrangian, so $\overline{m}$ is related to the quantity $m_R(\mu)$ via some renormalisation factors. The second term gives the lowest dimension vacuum condensate, where ${\cal C}\,=\,\left< \qq \right >$. This provides an excellent representation of all our solutions.
If we included the expression to all orders then the scales $\Lambda_1$
and $\Lambda_2$ would both be equal to $\Lambda_{QCD}$. However, the leading order forms in Eq.~(\ref{cond:eqn:opefit}) absorb different higher order contributions into the two terms and so $\Lambda_1$ and $\Lambda_2$ are in practice different, as we will discuss below. For large masses the condensate piece, $\cC$, is irrelevant and so it is the leading term that describes the mass function well. In contrast in the chiral limit, when $\overline{m}=0$, the second term of the OPE describes the behaviour of the mass function. This then accurately determines the scale $\Lambda_2$. Indeed, its value is equal to $\Lambda_{QCD}$. We can then in the chiral limit easily extract the renormalisation point independent condensate, $\cC\,\equiv\,\langle \qq \rangle$, from the asymptotics. 

For non-zero current masses,
 one can  attempt to fit both terms of the OPE in Eq.~(\ref{cond:eqn:opefit}) to the tail of the mass function, $M$ of Fig.~\ref{masses}. 
Comparing the full mass with $m_q \ne 0$ with that in the chiral limit, one sees how very small is the contribution that any condensate makes to the tail. So while a value for the condensate can be extracted, 
this procedure is not at all reliable. This is because of the difficulty in resolving the two functions in the OPE from one another and in fixing the appropriate scales, $\Lambda_1$ and $\Lambda_2$.

Strictly in the chiral limit, we may also extract the condensate using:
\begin{equation}\label{cond:eqn:tracecondensate}
\vspace{0.5mm}
	-\left<\qq\right>_\mu = Z_2\left(\mu,\Lambda\right) Z_m\left( \mu,\Lambda \right)N_c\, \tr_D \,\int^\Lambda \frac{d^4k}{\left( 2\pi \right)^4}\,S_F\left(k,\mu\right)\;,
\vspace{0.5mm}
\end{equation}
with $\left<\qq\right>_\mu$ the renormalisation point dependent quark condensate. At one-loop, this is related to the renormalisation point independent quark condensate:
\begin{equation}
\vspace{0.5mm}
	\left<\qq\right>_\mu = \left( \frac{1}{2}\log\frac{\mu^2}{\Lambda^2_{QCD}} \right)^{\gamma_m}\left<\qq\right>\;.
\vspace{0.5mm}
\end{equation}
which we compare with the asymptotic extraction to good agreement.

However, for non-zero quark masses, even for very small masses where the condensate is  still expected to play a sizeable role, we cannot apply Eq.~(\ref{cond:eqn:tracecondensate}), since it acquires a quadratic divergence, cf. Eq.~(\ref{cond:eqn:opefit}). Indeed, it is the elimination of this divergence that inspired the
definition proposed by Chang {\it et al.}~\cite{Chang:2006bm}, which, as we will mention below and explicitly show elsewhere, is unfortunately not equal to the condensate of the physical mass function and  is therefore
ambiguous. Consequently, we need a different definition, one close to the OPE, Eq.~(\ref{cond:eqn:opefit}).

This comes about by recalling that the Schwinger-Dyson equation, Eq.~(\ref{quark:sde}), has multiple solutions.
 In the chiral limit, there exist three solutions for $S_F(p)$ and its mass function $M(p^2)$. These correspond to the Wigner mode (the only solution accessible to perturbation theory), and two non-perturbative solutions of equal magnitude generated by the dynamical breaking of chiral symmetry. These we denote by $S_F^{\pm,W}(p)$, where:
\begin{equation}\label{cond:eqn:sols} 
\vspace{1mm}
	M(p^2) = 	\left\{ \begin{array}{l} 
				M^W(p^2) = 0\;\; \\ [-0.5mm]
                               \\[-0.5mm]
			      M^\pm(p^2) = \pm M^0(p^2) 
			\end{array} \right. \;.
\vspace{1mm}
\end{equation}
Analogous solutions to these exist as we move away from the chiral limit. The existence of these is essential to our ability to determine the $\qq$ condensate beyond the chiral limit, up to the strange quark mass.  While here we will present the results for the Maris-Tandy model, elsewhere we will provide the details for other more sophisticated modellings of QCD and for the extrapolation of lattice results to the continuum, which within the larger uncertainties inherent in these procedures confirm the results presented here.  

\begin{figure}[t]
	\centering
	{\includegraphics*[width=0.60\columnwidth]{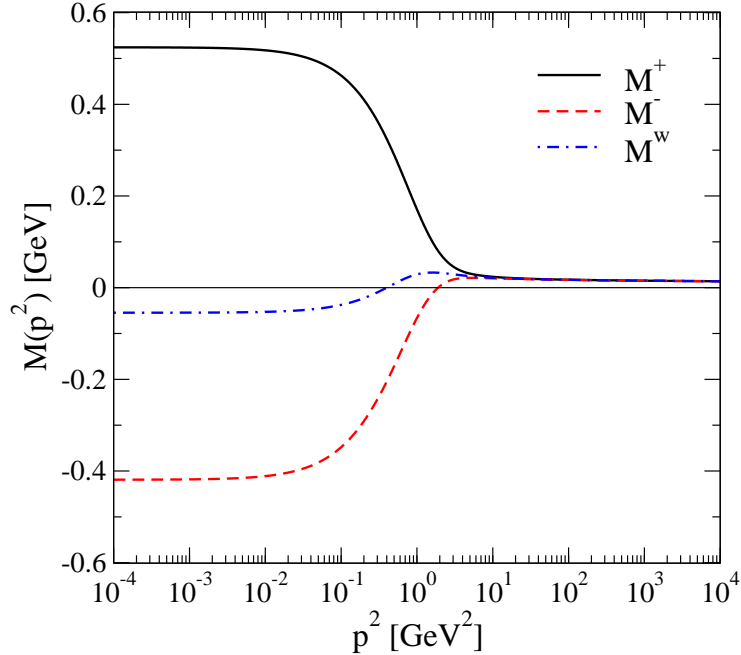}}
	\caption{Momentum dependence of the three solutions $M^\pm(p^2)$ and $M^W(p^2)$ for a quark mass $m(\mu)$=16 MeV, $\mu=19$ GeV.}\label{cond:fig:quark-sols}
\end{figure}

Crucially, for a given $m_R$ we obtain the $M^-$ and $M^W$ solutions by inserting the same $Z_2$ and $Z_m$ found for the $M^+$ solution. This ensures that differences in the dynamics of the three systems do not influence the ultraviolet running of the current-quark mass in the context of the subtractive renormalisation scheme used  by Maris and Tandy. The iteration process is performed using
Newton's method. For the solution $M^W$ this is mandatory, since it corresponds to a local maximum of the effective action and is therefore not accessible using the conventional fixed point iteration scheme. We renormalise at $\mu = 19$ GeV to compare with the work of Ref.~\cite{Chang:2006bm}. However, in the plots that follow we show the quark condensates as functions of the quark mass at $\mu =2$ GeV. This is obtained by one loop running from that at 19 GeV and just provides a convenient mass to plot. Later we will transform these momentum subtraction results  to the commonly used $\overline{MS}$ scheme at 2 GeV.
A representative example of the solutions is shown in Fig.~\ref{cond:fig:quark-sols}.

As we illustrate below the solutions $M^-$ and $M^W$ only exist below some critical mass $m_{cr}$ at which a bifurcation occurs. For zero active flavours and $\omega = 0.4$ GeV, we obtain $m_{cr} = 43.4$ MeV at $\mu=2$ GeV. The exact value of $m_{cr}$ is found to be model-dependent. Of course, the physical solution $M^+$ exists for all quark masses as shown in Fig.~\ref{masses}.
If one takes the difference between any pair of solutions of $M^{\pm},\,M^W$, then in each of these the first term in the OPE, Eq.(\ref{cond:eqn:opefit}), explicitly cancels and each has a behaviour qualitatively like $M^+$ in the chiral limit, Fig.~\ref{masses}, being dominated by a condensate term. We can then determine
this condensate, $\cal C$ in Eq.~(\ref{cond:eqn:opefit}), by integrating any of these differences, like $S_F^+-S_F^-$,
as in Eq.~(\ref{cond:eqn:tracecondensate}). Though such differences give a convergent integral, its value is just the difference of the condensates, for instance ${\cal C}^+-{\cal C}^-$ and this is not a measure of the physical condensate as Chang {\it et al.} presume. This is why we do not use a convergent integral to determine the physical condensate, but rather the OPE as we now describe.

\baselineskip=5.6mm
\section{Extracting the Condensate}

\begin{figure}[t]
	\centering
	{\includegraphics*[width=0.62\columnwidth]{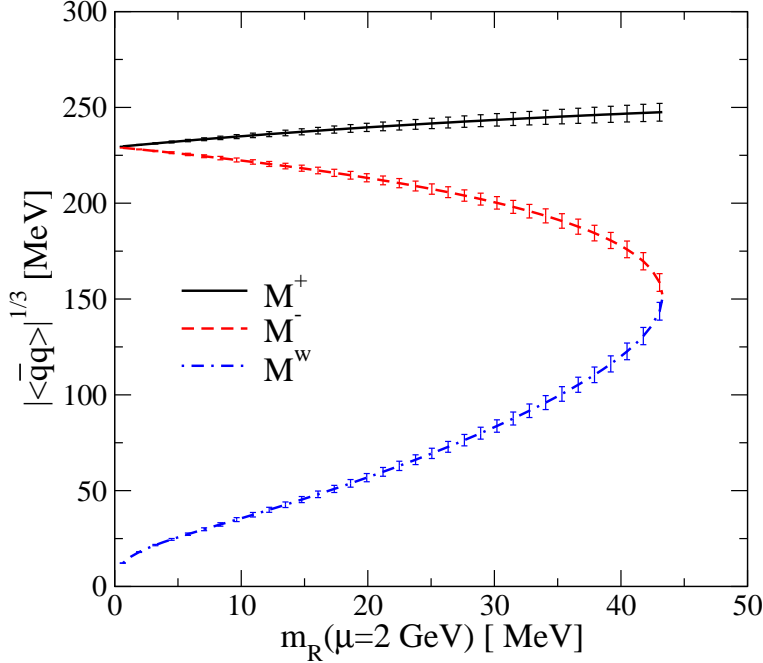}}
	\caption{Condensate extracted through simultaneous fitting of the three solutions to the fermion mass-function in the Maris-Tandy model with $N_f=0$ and $\omega=0.4$ GeV as functions of the current quark mass evolved to 2 GeV in a momentum subtraction scheme.}\label{cond:fig:cond-lm}
\end{figure}

\begin{figure}[ht]
  \centering
      {\includegraphics*[width=0.62\columnwidth]{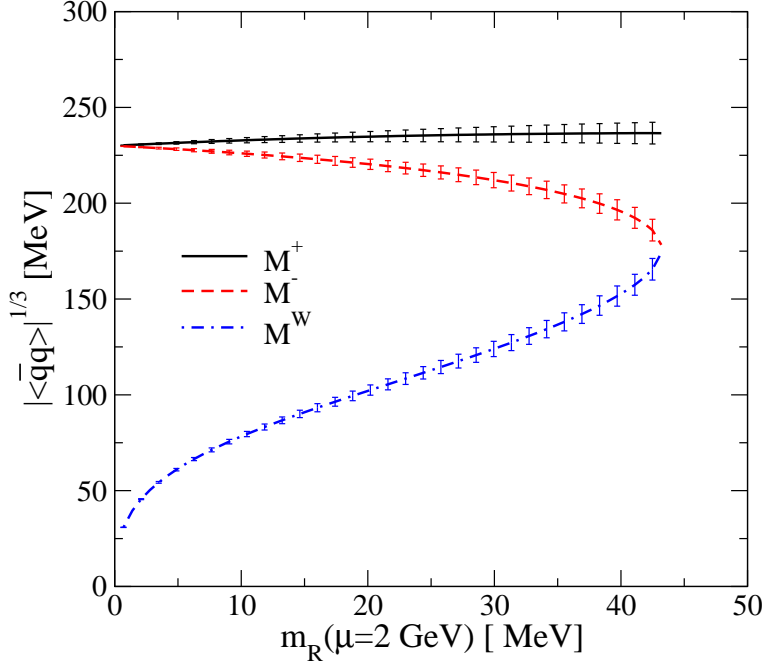}}
	\caption{Condensate for Maris-Tandy Model with $N_f=4$, $\omega=0.4$ GeV as a function of current quark mass evolved to 2~GeV, to be compared with Fig.~\ref{cond:fig:cond-lm}.}\label{cond:mt-4-4b}
\end{figure}

Instead of one single solution, we now have three solutions to the same model, each with identical running of the current-quark mass (the first term in Eq.~(\ref{cond:eqn:opefit})) in the ultraviolet region and differing only by their values of the condensate. Thus, for $m_R(\mu) < m_{cr}$,
it is possible to fit Eq.~(\ref{cond:eqn:opefit}) simultaneously to the three mass functions $M^\pm$, $M^W$. The scales $\Lambda_1$ and $\Lambda_2$ are 
found to be
the same for any current mass within the given model. $\Lambda_2$ is equal to $\Lambda_{QCD}$, while $\Lambda_1$ is roughly twice as big. Its exact value depends on the parameters of the model, like $\omega$ and $D$ of Eq.~(\ref{eqn:modelparam}). The condensates $\cC^\pm$ and $\cC^W$ are then determined
in an accurate and stable way. This fitting is performed using a modified Levenberg-Marquardt algorithm. The results are shown in  Figs.~\ref{cond:fig:cond-lm},~\ref{cond:mt-4-4b} for $N_f=0, 4$. The error bars reflect the accuracy
with which the mass functions  represented by two terms in the OPE expression, Eq.~(\ref{cond:eqn:opefit}), are separable with the anomalous dimensions specified.

We see that within errors the condensate is found to increase with quark mass.
This rise at small masses was anticipated by Novikov {\it et al.}~\cite{Novikov:1981xj} combining a perturbative chiral expansion with QCD sum-rule arguments. That the chiral logs relevant at very small $m_q$ are barely seen is due to the quenching of the gluon and the rainbow approximation of Eq.~(\ref{cond:eqn:quarksde1}). As we will show elsewhere this has little effect when we model more complex interactions including matching with the lattice. 

We see in Figs.~\ref{cond:fig:cond-lm}, \ref{cond:mt-4-4b} that the $M^-$ and $M^W$ solutions bifurcate below $m_{cr} \simeq 43.4 (44.0)$ MeV with $\omega =0.4$ GeV for $N_f=0(4)$ respectively.
 But what about the value of the condensate for the physical solution $M^+$ beyond the region where $M^-$ and $M^W$ exist, {\it i.e.} $m_R(\mu) > m_{cr}$?
Having accurately determined the scales $\Lambda_1$ and $\Lambda_2$ in the OPE of Eq.~(\ref{cond:eqn:opefit}) in the region where all 3 solutions exist, we
could
 just continue to use the same values in fitting the physical $M^+$ solution alone and find its condensate.
However, this would make
it difficult to produce realistic errors as the quark mass increases.

\begin{figure}[p]
\centering      
     {\includegraphics*[width=0.62\columnwidth]{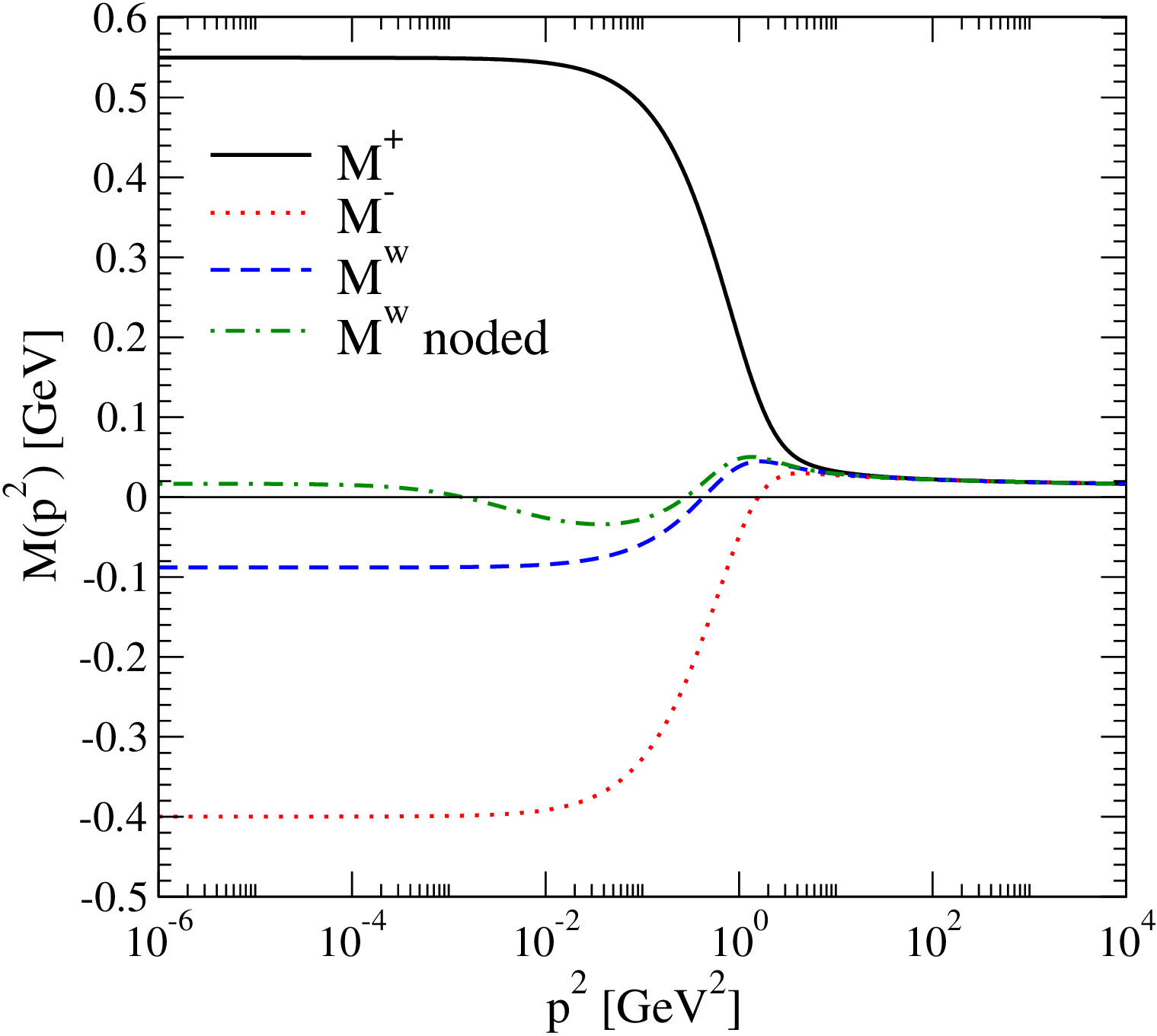}}
   \caption{Momentum dependence of the 4 solutions for the fermion
mass-function in the Maris-Tandy model with $m=20$ MeV at $\mu=19$ GeV,
$n_F$=4, $\omega=0.4$ GeV.}\label{cond:noded}
\vspace{4mm}
\centering     
     {\includegraphics*[width=0.62\columnwidth]{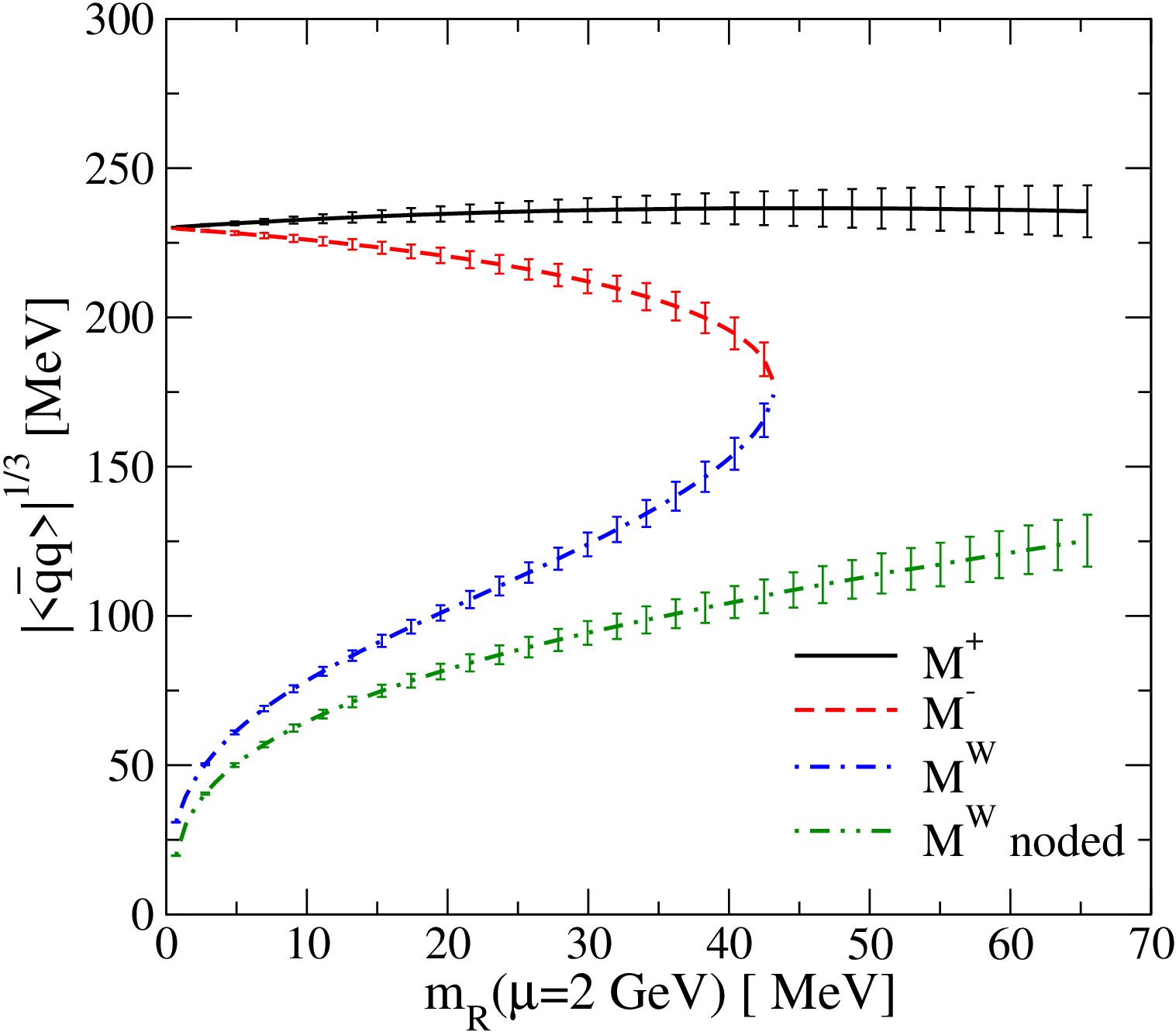}}
   \caption{Current quark mass dependence of the condensates for Maris-Tandy
model with $n_F=4$, $\omega=0.4$ GeV, including the noded solution of
Fig.~\ref{cond:noded}.}\label{cond:nodecond}
\end{figure}

As soon as one allows for solutions for the fermion mass-function that are not positive definite, one exposes a whole series of variants on the solutions $M^-$, $M^W$ we have already considered. Thus there are {\it noded} solutions, which have also been discovered recently in the context of a simple Yukawa theory  by Martin and Llanes-Estrada~\cite{Martin:2006qd}. We illustrate this within the Maris-Tandy model, for instance with $N_f=4$ and $\omega=0.4$, in Fig.~\ref{cond:noded}. There the four solutions we have found are displayed.
\begin{figure}[t]
  \centering
      {\includegraphics*[width=0.62\columnwidth]{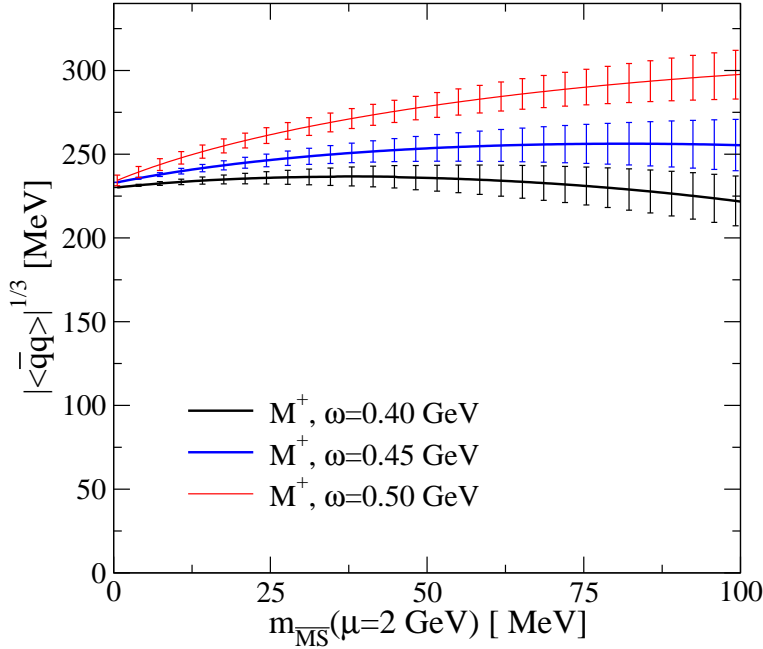}}
	\caption{Condensate for Maris-Tandy Model with $N_f=4$, $\omega=0.4,0.45,0.5$ GeV as a function of current quark mass defined at 2~GeV in $\overline{MS}$ scheme.}\label{cond:m45MS}
\end{figure} 
It is interesting to note that this noded solution is not limited to the same domain that restricts $M^-$ and $M^W$. These noded solutions do develop a singularity in $M(p^2)$ beyond $m=51.4$ MeV at $\mu=2$ GeV. However, this is compensated for by a zero in $\cA(p^2)$, Eq.~(\ref{fermion:eqn:AB}), until $m=66.3$ MeV.  Thus there exists a solution with a well-defined ultraviolet running of the quark mass exactly as the $M^+$ solution, as far as $m=66.3$ MeV. While at small quark masses we have all four solutions, at larger masses there are still two. Consequently, we can accurately fix
the scales $\Lambda_1$ and $\Lambda_2$ of Eq.~(\ref{cond:eqn:opefit}) at each $m_{cr}$, and so determine the condensates as shown in Fig.~\ref{cond:nodecond}.
Indeed, fitting the $M^+$ and $M^W_{noded}$ at each value of $m_R(\mu)$ with common scales in the OPE equation, Eq.~(\ref{cond:eqn:opefit}) allows the condensate for the physical solution to be found for much larger quark masses, as shown in Fig.~\ref{cond:nodecond}. Indeed, these fits confirm that $\Lambda_1$ and $\Lambda_2$ are independent of $m_R(\mu)$. We can then fit the remaining $M^+$ solutions shown in Fig.~\ref{masses} to give the physical condensate shown in Fig.~\ref{cond:nodecond} for acceptable values of $\omega$ as determined by ~\cite{Alkofer:2002bp}. In Fig.~\ref{cond:m45MS} we have in fact scaled the quark mass from $\mu = 2$ GeV in the (quark-gluon) MOM scheme by one loop running to the
$\overline{MS}$ scheme at 2 GeV using the relationship between $\Lambda_{MOM}$ and $\Lambda_{\overline{MS}}$ for 4 flavours deduced by Celmaster and Gonsalves~\cite{Celmaster:1979km}. In this latter scheme the strange quark mass is $\sim 95$ MeV~\cite{PDG}.

Within the range of the Maris-Tandy modelling of strong coupling QCD we find that the ratio of the condensates at the strange quark mass to the chiral limit is
\begin{equation}
	\left<\qq\right>_{m(\overline{MS})\,=95\,{\rm MeV}}/\left<\qq\right>_{m=0}\;=\;(\,1.1\,\pm\,0.2\,)^3.
\end{equation}
in a world with 4 independent flavours. In a longer paper we will consider
the Schwinger-Dyson equations with more sophisticated ans\"atze incorporating the quark, gluon and ghost interactions both in the continuum and on the lattice.
Moreover, here all the quarks have the same mass and there is no mixing between  different hidden flavour pairs. Elsewhere we will illustrate the change that occurs in solving the quark Schwinger-Dyson equations with 2 flavours
of very small mass $m_{u,d}$ and 1 flavour with variable mass. Of course, in the quenched case quark loops decouple and exactly replicate the results given here.  
Chang {\it et al.} have proposed that the fact that the $M^-$ and $M^W$ solutions only exist for some domain of quark masses, $m_q\,\in\,[0,m_{cr}]$, is directly linked to the domain of convergence of the chiral expansion~\cite{Hatsuda:1990tt,Meissner:1994wy}. While the existence of multiple solutions for the fermion mass function is essential for the extraction of the condensate, we note that distinct domains exist for
the different solutions and that the simplest noded solution exists in a larger domain. Though the value of $m_{cr}$ is indeed some measure of the
range of validity of the chiral expansion, the fact that $m_{cr}$
is both strongly model and solution dependent indicates its value is no more than a guide and not likely to be the exact bound claimed by Chang {\it et al.}

What we have shown here is that there is a robust method of determining the value of the $\qq$ condensate beyond the chiral limit based on the Operator Product Expansion. Of course, as the
quark mass increases the contribution of the condensate to the behaviour of the mass function, Fig.~\ref{masses}, becomes relatively lass important and so the errors on the extraction of the physical condensate increases considerably.
Nevertheless, the method is reliable up to and beyond the strange quark mass.
Alternative definitions are not.
\vspace{1cm}

\centerline{\bf Acknowledgments}

RW is grateful to the UK Particle Physics and Astronomy Research Council (PPARC) for the award of a research studentship.
We thank Roman Zwicky and Dominik Nickel for interesting discussions.
This work
was supported in part by the EU Contract No. MRTN-CT-2006-035482, 
\lq\lq FLAVIAnet''.

\end{document}